\documentclass[a4paper,10pt,oneside]{article}
\usepackage[T1]{fontenc}
\usepackage{amsmath}
\usepackage{amssymb}
\usepackage{graphicx}
\usepackage{cite}

\newcommand{\Tr}{\mathop{\mathrm{Tr}}\nolimits}
\newcommand{\Int}[4]{\int\nolimits_{#1}^{#2}{#3}\,\mathrm{d}{#4}}
\newcommand{\Ints}[3]{\int\nolimits_{#1}^{#2}\mathrm{d}{#3}\,}
\newcommand{\sumi}{\sum_{k=1}^{3V}}
\newcommand{\li}{\lambda_k}
\newcommand{\lis}{{\lambda_k^*}}
\newcommand{\sgn}{\mathop{\mathrm{sgn}}\nolimits}
\newcommand{\figheight}{60mm}

\newcommand{\institute}[1]{\parbox{16cm}{\begin{center}\normalsize
    \sl #1 \end{center}}}

\begin{document}

\title{{\normalsize\vspace*{-2cm}\hfill\mbox{ITP-Budapest 632}\\\vspace*{2cm}}
Hadron spectroscopy from canonical partition functions}
\author{Z.~Fodor$^{a,b,c}$, K.~K.~Szab\'o$^{a}$ and
B.~C.~T\'oth$^{a,b}$\\ \vspace{1mm} 
\institute{$^a$Department of Physics, University of Wuppertal, Germany\\
$^b$Institute for Theoretical Physics, E\"otv\"os University, Budapest,
Hungary\\
$^c$Department of Physics, University of California, San Diego,
 USA}}

\date{\today}

\maketitle

\begin{abstract}
  A spectroscopic method for staggered fermions based on
  thermodynamical considerations is proposed. The canonical partition
  functions corresponding to the different quark number sectors are
  expressed in the low temperature limit as polynomials of the
  eigenvalues of the reduced fermion matrix. Taking the zero
  temperature limit yields the masses of the lowest states. The method
  is successfully applied to the Goldstone pion and both dynamical and
  quenched results are presented showing good agreement with that of
  standard spectroscopy. Though in principle the method can be used to
  obtain the baryon and dibaryon masses, due to their high
  computational costs such calculations are practically unreachable.
\end{abstract}

\section{Introduction}

One of the most critical steps in hadron spectroscopy is the choice of the
wave function for the particle of interest. Without finding the proper
interpolating operator contamination from other particle states can often
occur. This does not cause serious difficulties when one deals with simple
particles such as mesons or baryons, but the extraction of the correct mass
can be a tremendous task when one is looking for more compound
states. Therefore, a spectroscopic method that does not require knowledge of
the wave function is desirable.

The question was first addressed in Ref.~\cite{Gibbs:1986hi}, where the
explicit inversion of the fermion matrix on duplicated lattice configurations
and the examination of the exponential decay rate of the lowest hadron state
led to the expression
\begin{equation}
  am_{\pi} = -\frac{1}{L_t} \cdot \max_{\left| \lambda_k \right| < 1} \ln
  \left| \lambda_k \right|^2
  \label{eqn:gibbs}
\end{equation}
for the mass of the Goldstone pion on each configuration, where $a$ is the
lattice spacing, $L_t$ is the number of lattice sites in the temporal
direction and $\lambda_k$ are the eigenvalues of the reduced staggered fermion
matrix (see Section \ref{sec:cpf_lattice}). The aim is to find the relation
between the hadron spectrum and these eigenvalues.

There has been many advances recently in the canonical approach to finite
density QCD \cite{Alexandru:2005ix, Kratochvila:2005mk, Kratochvila:2006jx,
deForcrand:2006ec}. Based on the canonical formulation we make an attempt to
clarify and extend the findings of Ref.~\cite{Gibbs:1986hi} and give a method
which in principle can be used to obtain the masses of different particles.

The paper is organized as follows. In Section \ref{sec:cpf_mass} we summarize
how canonical partition functions can be used to obtain the masses of
particles. This is followed by a description of how canonical partition
functions can be obtained on the lattice in Section \ref{sec:cpf_lattice}. The
way one can find the relevant eigenvalues is shown in section
\ref{sec:eigenvalues}. In Section \ref{sec:z3} we show how the $Z_3$
symmetry can be used to simplify our formulae. The case of baryons is
explained in Section \ref{sec:baryon} while the case of mesons is discussed in
Section \ref{sec:isospin}. Finally, after showing our numerical results in
Section \ref{sec:results} we conclude in Section \ref{sec:conclusions}.

\section{Masses from canonical partition functions}
\label{sec:cpf_mass}

Let us consider the general case when we have $n_s$ different quark fields.
Let $\hat{N}_i$ and $\mu_i$ denote the quark number operator and the quark
number chemical potential corresponding to the $i$th quark field,
respectively. Then the grand canonical partition function at a given set of
chemical potential values $(\mu_1,\mu_2,\dots,\mu_{n_s})$ and temperature $T$
is given by
\begin{equation}
Z(\mu_1,\mu_2,\dots,\mu_{n_s},T) = \Tr \left[ e^{-(\hat{H} -\mu_1\hat{N}_1
  -\mu_2\hat{N}_2-\, \dots\, - \mu_{n_s}\hat{N}_{n_s})/T} \right].
\end{equation}
The canonical partition function corresponding to a given set of quark number
values $N_1,\dots,N_{n_s}$ can be obtained by taking the trace only over the
subspace $\hat{N}_1=N_1, \dots, \hat{N}_{n_s}=N_{n_s}$.
\begin{equation}
\begin{split}
  Z_{N_1,\dots,N_{n_s}}(T) &= \Tr\left[
    e^{-\hat{H}/T}\cdot\delta_{\hat{N}_1,N_1}\,
    \dots \, \delta_{\hat{N}_{n_s},N_{n_s}} \right] \\
  &= \Tr\left[e^{-\hat{H}/T}\cdot\frac{1}{2\pi}\Int{0}{2\pi}{e^{i(\hat{N}_1-
        N_1)\theta_1}}{\theta_1}\,\cdots\,
    \frac{1}{2\pi}\Int{0}{2\pi}{e^{i(\hat{N}_{n_s}-
        N_{n_s})\theta_{n_s}}}{\theta_{n_s}} \right] \\
  &= \frac{1}{(2\pi)^{n_s}} \, \Ints{0}{2\pi}{\theta_1}
  e^{-iN_1\theta_1}\,\dots
  \Ints{0}{2\pi}{\theta_{n_s}} e^{-iN_{n_s}\theta_{n_s}}\,
  \Tr \left[e^{-(\hat{H} -iT\theta_1\hat{N}_1
      -\,\dots\,- iT\theta_{n_s}\hat{N}_{n_s})/T}\right] \\
  &= \frac{1}{(2\pi)^{n_s}} \, \Ints{0}{2\pi}{\theta_1}
  e^{-iN_1\theta_1}\,\dots 
  \Ints{0}{2\pi}{\theta_{n_s}} e^{-iN_{n_s}\theta_{n_s}}\,
  Z(iT\theta_1,\dots,iT\theta_{n_s},T)  
\end{split}
\end{equation}
When one introduces imaginary chemical potentials \cite{Roberge:1986mm}, the
different canonical partition functions become the coefficients in the Fourier
expansion of the grand canonical partition function.
\begin{equation}
Z_{N_1,\dots,N_{n_s}}(T) = \frac{1}{(2\pi T)^{n_s}} \, \Ints{0}{2\pi T}{\mu_1}
\dots \Ints{0}{2\pi T}{\mu_{n_s}} e^{-i\mu_1 N_1/T} \cdots e^{-i\mu_{n_s}
  N_{n_s}/T} \, Z(i\mu_1,\dots,i\mu_{n_s},T)
\end{equation}
\begin{equation}
  Z(i\mu_1,\dots,i\mu_{n_s},T)= \sum_{N_1=-\infty}^{\infty} \cdots
  \sum_{N_{n_s}= 
    -\infty}^{\infty} Z_{N_1,\dots,N_{n_s}}(T) \, e^{i\mu_1 N_1/T} \cdots
  e^{i\mu_{n_s} N_{n_s}/T}
\end{equation}

When the aim is to find the energy of the lowest state in the sector
corresponding to quark numbers $(N_1,\dots,N_{n_s})$ one has to examine the
low temperature behavior of the free energy
\begin{equation}
  F_{N_1,\dots,N_{n_s}}(T) = - T \ln Z_{N_1,\dots,N_{n_s}}(T).
\end{equation}
The canonical partition function can be written as
\begin{equation}
Z_{N_1,\dots,N_{n_s}}(T) = \sum_{k=0}^{\infty} n_k^{(N_1,\dots,N_{n_s})} \,
  e^{-E_k^{(N_1,\dots,N_{n_s})}/T},
\end{equation}
where $E_k^{(N_1,\dots,N_{n_s})}$ and $n_k^{(N_1,\dots,N_{n_s})}$ are the
energy and the multiplicity of the $k$th state in sector
$(N_1,\dots,N_{n_s})$, respectively. In sector $(0,\dots,0)$ the lowest state
is the vacuum, which is assumed to be non-degenerate.
\begin{equation}
  Z_{0,\dots,0}(T) = e^{-E_0^{(0,\dots,0)}/T} + \sum_{k=1}^{\infty}
  n_k^{(0,\dots,0)} \,  e^{-E_k^{(0,\dots,0)}/T}
\end{equation}
Then the difference of the free energies of sector $(N_1,\dots,N_{n_s})$ and
the zero quark number sector can be rewritten as
\begin{multline}
F_{N_1,\dots,N_{n_s}}(T) - F_{0,\dots,0}(T) = E_0^{(N_1,\dots,N_{n_s})} -
E_0^{(0,\dots,0)} - T \ln n_0^{(N_1,\dots,N_{n_s})} \\
- T \ln \left[   \frac{ \displaystyle 1 +
    \sum_{k=1}^{\infty}
    \frac{n_k^{(N_1,\dots,N_{n_s})}}{n_0^{(N_1,\dots,N_{n_s})}} \,
    e^{-(E_k^{(N_1,\dots,N_{n_s})} - E_0^{(N_1,\dots,N_{n_s})})/T}}
  {\displaystyle 1 +
    \sum_{k=1}^{\infty}
    n_k^{(0,\dots,0)}\, e^{-(E_k^{(0,\dots,0)} - E_0^{(0,\dots,0)})/T}}
\right].
\label{eqn:fdifflong}
\end{multline}
The mass of the lowest state in sector $(N_1,\dots,N_{n_s})$ is the difference
of the energy of the ground state in this sector and the energy of the vacuum
state,
\begin{equation}
m_0^{(N_1,\dots,N_{n_s})} = E_0^{(N_1,\dots,N_{n_s})} - E_0^{(0,\dots,0)}.
\end{equation}

If the temperature is much smaller than the energy differences that appear in
the exponentials in equation (\ref{eqn:fdifflong}) then the last term on the
r.h.s.\ of (\ref{eqn:fdifflong}) is negligible compared to the other terms. In
this region the difference of the free energies follows a linear behaviour,
\begin{equation}
  F_{N_1,\dots,N_{n_s}}(T) - F_{0,\dots,0}(T)  \approx
  m_0^{(N_1,\dots,N_{n_s})} - T \ln  n_0^{(N_1,\dots,N_{n_s})},
\label{eqn:fdifflin}
\end{equation}
where the slope of the linear behaviour depends only on the multiplicity of
the ground state. Therefore, the mass of the lightest particle carrying
quantum numbers $(N_1,\dots,N_{n_s})$ and its multiplicity can be obtained by
a linear extrapolation to the $T=0$ limit.
\begin{equation}
m_0^{(N_1,\dots,N_{n_s})} =
\lim_{T\to 0} \left[ F_{N_1,\dots,N_{n_s}}(T) - F_{0,\dots,0}(T) \right]
\label{eqn:fdiff3}
\end{equation}

\section{Canonical partition functions on the lattice}
\label{sec:cpf_lattice}

The temperature on the lattice is given by $T=1/aL_t$, where $L_t$ is the
number of sites in the temporal direction and $a$ is the lattice spacing. Let
$\hat{\mu}_i=\mu_ia$ denote the chemical potentials in lattice units. In order
to introduce these chemical potentials on the lattice the forward time-like
links have to be multiplied by $e^{i\hat{\mu}_i}$ and the backward time-like
links by $e^{-i\hat{\mu}_i}$ in the fermion determinant of quarks of type $i$
\cite{Hasenfratz:1983ba}. Then the grand canonical partition function using
staggered lattice fermions can be written as
\begin{equation}
  Z(i\hat{\mu}_1,\dots,i\hat{\mu}_{n_s}) = \int[\mathrm{d}U]\,e^{-S_g[U]}\,
  \prod_{i=1}^{n_s} \det
  M(m_i,i\hat{\mu}_i,U)^{n_i/4},
\label{eqn:zmulat1}
\end{equation}
where $m_i$ denotes the bare mass and $n_i$ denotes the number of tastes of
the $i$th staggered quark field. The functional integral is taken over all
possible gauge configurations $U$ and $S_g[U]$ denotes the pure gauge part of
the action.

The partition function can be rewritten in the form
\begin{equation}
  Z(i\hat{\mu}_1,\dots,i\hat{\mu}_{n_s}) = \int[\mathrm{d}U]\,e^{-S_g[U]}\,
  \prod_{i=1}^{n_s} \det M(m_i,0,U)^{n_i/4} \times \prod_{i=1}^{n_s} \left(
  \frac{\det M(m_i,i\hat{\mu}_i,U)}{\det M(m_i,0,U)} \right)^{n_i/4}.
\end{equation}
The ratios of the determinants can be treated as observables while the
functional integral can be taken using the measure at $\hat{\mu}_i=0$.
Then the partition function becomes the expectation value of
the determinant ratios taken over the ensemble generated at zero chemical
potentials,
\begin{equation}
  Z(i\hat{\mu}_1,\dots,i\hat{\mu}_{n_s}) 
  = Z \cdot \left<  \prod_{i=1}^{n_s} \left( 
  \frac{\det M(m_i,i\hat{\mu}_i,U)}{\det M(m_i,0,U)}
  \right)^{n_i/4}  \right>,
  \label{eqn:zmulat3}
\end{equation}
where $Z$ denotes the zero chemical potential value of the partition function
\cite{Fodor:2001pe}.

Therefore, the canonical partition functions are obtained by taking the
expectation values of the Fourier components of the determinant ratios.

\newlength{\ZNN}
\begin{equation}
  Z_{N_1,\dots,N_{n_s}} =  Z \cdot
  \Bigg< \prod_{i=1}^{n_s}
  \frac{L_t}{2\pi} \Ints{0}{\frac{2\pi}{L_t}}{\hat{\mu}_i}
  e^{-i\hat{\mu}_i N_i L_t} \, \left( \frac{\det M(m_i,i\hat{\mu}_i,U)}{\det
    M(m_i,0,U)} \right)^{n_i/4} \Bigg>
\label{eqn:canonlat1}
\end{equation}

In order to perform the assigned Fourier transformations we need the analytic
$\hat{\mu}$--dependence of $\det M(i\hat{\mu})$. In temporal gauge, the fermion
matrix can be written as
\begin{equation}
M(i\hat{\mu})= \begin{pmatrix} B_0 & e^{i\hat{\mu}} & 0 & \dots & 0 &
  Ue^{-i\hat{\mu}}  \\
-e^{-i\hat{\mu}} & B_1 & e^{i\hat{\mu}} & \dots & 0 & 0 \\
0 & -e^{-i\hat{\mu}} & B_2 & \dots & 0 & 0 \\
\vdots & \vdots & \vdots & \ddots & \vdots & \vdots \\
0 & 0 & 0 & \dots & B_{L_t-2} & e^{i\hat{\mu}} \\
-U^\dagger e^{i\hat{\mu}} & 0 & 0 & \dots & -e^{-i\hat{\mu}} & B_{L_t-1}
\end{pmatrix},
\label{eqn:ferm_matrix}
\end{equation}
where U denotes the remaining time direction links on the last timeslice
(including the correct staggered phases) and $B_k$ is the spacelike staggered
fermion matrix on timeslice $k$. In matrix (\ref{eqn:ferm_matrix}) each block
is a $3V \times 3V$ matrix, where $V=L_s^3$ and $L_s$ is the spatial size of
the lattice. After performing $L_t-2$ steps of Gaussian elimination, the
determinant of (\ref{eqn:ferm_matrix}) can be written as
\begin{equation}
  \det M(i\hat{\mu})= e^{3VL_ti\hat{\mu}} \det \left(S-e^{-i\hat{\mu}
      L_t}\right), 
\label{eqn:det1}
\end{equation}
where
\begin{equation}
S= \begin{pmatrix} 0 & 1 \\ 1 & B_{L_t-1} \end{pmatrix}
\begin{pmatrix} 0 & 1 \\ 1 & B_{L_t-2} \end{pmatrix} \cdots
\begin{pmatrix} 0 & 1 \\ 1 & B_0 \end{pmatrix}
\begin{pmatrix} U & 0 \\ 0 & U \end{pmatrix}
\label{eqn:S}
\end{equation}
is the $6V\times 6V$ sized reduced fermion matrix \cite{Hasenfratz:1991ax}.
Let $\lambda_k$ denote the eigenvalues of $S$. Then (\ref{eqn:det1}) can be
written as
\begin{equation}
\det M(i\hat{\mu})= e^{3VL_ti\hat{\mu}} \prod_{k=1}^{6V} \left( \lambda_k
  -e^{-i\hat{\mu} L_t}\right),
\end{equation}
and thus,
\begin{equation}
  \frac{\det M(i\hat{\mu})}{\det M(0)} =  e^{3VL_ti\hat{\mu}} \prod_{k=1}^{6V}
  \frac{\lambda_k -e^{-i\hat{\mu} L_t}}{\lambda_k -1}.
\label{eqn:detdet1}
\end{equation}

The eigenvalues of matrix $S$ have a symmetry, according to which whenever
$\lambda$ is an eigenvalue of $S$ then $1/\lambda^*$ is also an eigenvalue of
$S$ \cite{Hasenfratz:1991ax}. Therefore, each eigenvalue whose absolute value
is greater than 1 has a pair with an absolute value smaller than 1, and vice
versa. (We will not deal with the case when at least one of the eigenvalues
lie on the unit circle because these gauge configurations constitute a zero
measure set.) Then (\ref{eqn:detdet1}) can be written as
\begin{equation}
  \frac{\det M(i\hat{\mu})}{\det M(0)} = e^{3VL_ti\hat{\mu}} \prod_{k=1}^{3V}
  \frac{\lambda_k -e^{-i\hat{\mu} L_t}}{\lambda_k -1} \, 
  \frac{\frac{1}{\lambda_k^*} -e^{-i\hat{\mu} L_t}}{\frac{1}{\lambda_k^*} -1}
  =   \prod_{k=1}^{3V} \left| \frac{1 - \lambda_k \,e^{i\hat{\mu} L_t}}{ 1 -
      \lambda_k}  \right|^2,
\label{eqn:detdet2}
\end{equation}
where the product is taken over only the eigenvalues lying inside the unit
circle. From now on when the limits of a sum or product taken over the
eigenvalues of $S$ are from 1 to $3V$, then the sum or product is meant to be
taken over only the ``small'' eigenvalues, that is, the eigenvalues with
absolute value smaller than 1.

\begin{figure}
\begin{center}
\resizebox{!}{\figheight}{\includegraphics{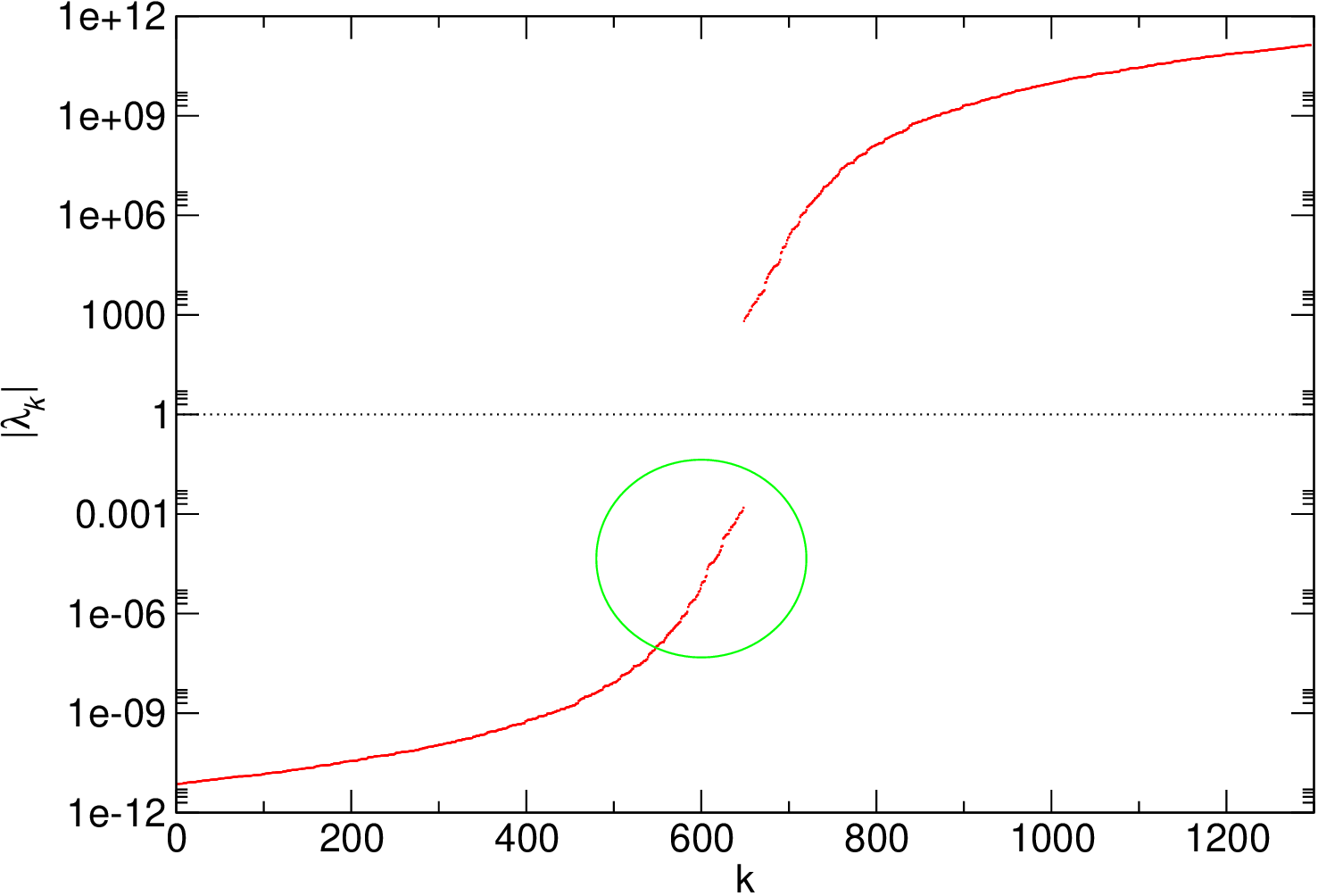}}
\end{center}
\caption{The absolute values of the eigenvalues of the reduced fermion matrix
  $S$ on a typical $6^3\times 24$ lattice configuration at a temperature of
  $T\approx 25\,\text{MeV}$. At this temperature there is already a noticeable
  gap between the eigenvalues lying inside the unit circle and the ones lying
  outside. The eigenvalues that are relevant for calculating the canonical
  partition functions are circled.}
\label{fig:gap}
\end{figure}

When the temperature is low ($T \ll T_c$) a gap appears between the ``small''
and ``large'' eigenvalues of $S$ (see Figure \ref{fig:gap}).  This makes a
Taylor expansion of (\ref{eqn:detdet2}) in the small eigenvalues possible. As
the temperature decreases the small eigenvalues become exponentially smaller,
increasing the validity of the series expansion.  Including the rational
exponent for the number of tastes $n_t$ a first order expansion gives

\begin{equation}
\begin{split}
  \left( \frac{\det M(i\hat{\mu})}{\det M(0)} \right)^{n_t/4} &= \left(
    \prod_{k=1}^{3V} \left| \frac{1 - \lambda_k \,e^{i\hat{\mu} L_t}}{ 1 -
        \lambda_k} \right|^2 \right)^{n_t/4} \\
  &\approx \left[ 1 + \frac{n_t}{4} \sumi\li + \frac{n_t}{4} \sumi\lis \right]
  + e^{i\hat{\mu} L_t} \left[ -\frac{n_t}{4} \sumi\li \right] + e^{-i\hat{\mu}
    L_t} \left[ -\frac{n_t}{4} \sumi\lis \right].
\label{eqn:1stexp}
\end{split}
\end{equation}
By performing an $n$th order Taylor expansion we explicitly obtain all the
Fourier coefficients up to $n$th order.

This way the assigned Fourier transformations in equation
(\ref{eqn:canonlat1}) can be performed easily configuration by configuration
by simply choosing the coefficients of the corresponding exponential
terms. The order of the leading order term for sector $(N_1,\dots,N_{n_s})$ is
$\left| N_1 \right | + \dots + \left| N_{n_s} \right|$. When all the quark
fields have 4 tastes ($n_i=4$) the leading order term for sector
$(N_1,\dots,N_{n_s})$ can be written as
\begin{equation}
  Z_{N_1,\dots,N_{n_s}} \stackrel{\text{LO}}{=} Z \cdot \left<
    \prod_{i=1}^{n_s} 
    \left[ (-1)^{\left|N_i \right|} \sum_{1 \le k_1^{(i)} < \dots < k_{\left|
            N_i \right|}^{(i)} \le 3V} 
      \left(  \lambda_{k_1^{(i)}}^{(i)} \cdots \lambda_{k_{\left| N_i
            \right|}^{(i)}}^{(i)} \right)^{*(\sgn N_i)} \right] \right>,
\label{eqn:znlo4}
\end{equation}
where $*(\sgn N_i)$ in the exponent means that there is a complex conjugation
if $N_i$ is negative. $\lambda_k^{(i)}$ stands for the $k$th eigenvalue of the
reduced matrix $S^{(i)}$ obtained from the fermion matrix of the $i$th quark
field.

The leading order term in case of arbitrary number of tastes $n_i$ can be
obtained from (\ref{eqn:znlo4}) as follows. The formula within the expectation
value signs can be written as a homogeneous polynomial of the eigenvalues of
degree $\left| N_1 \right | + \dots + \left| N_{n_s} \right|$ using the
expressions
\begin{equation}
\begin{split}
\sum_{k=1}^{3V} \left(\lambda_k^{(i)}\right)^{j} \qquad  j=1,\dots,N_i \qquad
\text{if $N_i$ is positive, and}\\
\sum_{k=1}^{3V} \left({\lambda_k^{(i)}}^*\right)^{j} \qquad  j=1,\dots,-N_i
\qquad \text{if $N_i$ is negative.}
\label{eqn:expvar}
\end{split}
\end{equation}
The leading order term in the general case is obtained by replacing the
expressions 
\begin{equation}
\sum_{k=1}^{3V} \left(\lambda_k^{(i)}\right)^{j} \quad \text{with}\quad
\frac{n_i}{4}\sum_{k=1}^{3V} \left(\lambda_k^{(i)}\right)^{j} 
\qquad \text{and}\qquad
\sum_{k=1}^{3V} \left({\lambda_k^{(i)}}^*\right)^{j}\quad \text{with}\quad
\frac{n_i}{4}\sum_{k=1}^{3V} \left({\lambda_k^{(i)}}^*\right)^{j}
\label{eqn:expreplace}
\end{equation}
in the above polynomial.

\section{Obtaining the relevant eigenvalues}
\label{sec:eigenvalues}

In order to calculate the canonical partition functions using the description
given in (\ref{eqn:znlo4}), (\ref{eqn:expvar}) and (\ref{eqn:expreplace}) we
do not need all of the $3V$ small eigenvalues. At lower temperatures ($L_t \ge
50-100$) the small eigenvalues alone span a range of 20--40 orders of
magnitude. Therefore, the relevant eigenvalues that contribute significantly
to the sums in (\ref{eqn:znlo4})--(\ref{eqn:expreplace}) are the largest few
of the small eigenvalues (see Figure \ref{fig:gap}). Since the condition
number of matrix $S$ at low temperatures can be in the range of
$O(10^{60}-10^{100})$ and these relevant eigenvalues are in the middle of the
spectrum, finding these eigenvalues seems practically impossible.

Nevertheless, the matrix $S$ has some symmetry properties that make it
possible. The spacelike staggered fermion matrices $B_k$, which appear in
(\ref{eqn:ferm_matrix}) and (\ref{eqn:S}), obey a $\gamma_5$-hermiticity
\begin{equation}
\gamma_5 B_k = B_k^{\dagger} \gamma_5,
\end{equation}
where
\begin{equation}
\left(\gamma_5\right)_{xy} = \delta_{xy} \cdot (-1)^{\sum_\mu x_\mu}.
\end{equation}
Therefore, the inverse of $S$ can be obtained as
\begin{equation}
S^{-1} = (-1)^{L_t+1} 
\begin{pmatrix} 0 & -\gamma_5 \\ \gamma_5 & 0 \end{pmatrix}
S^{\dagger}
\begin{pmatrix} 0 & -\gamma_5 \\ \gamma_5 & 0 \end{pmatrix}.
\label{eqn:Sinv}
\end{equation}

As a consequence, once we have the matrix $S$ both $S+S^{-1}$ and $S-S^{-1}$
can be easily constructed. Then by inverting these two one can arrive at
\begin{equation}
Q = \frac12 \big[ \left(S+S^{-1}\right)^{-1} - \left(S-S^{-1}\right)^{-1}
\big].
\end{equation}
The order of magnitude of the condition number of $S+S^{-1}$ and $S-S^{-1}$ is
less than half of that of $S$. Therefore, much less numerical precision is
sufficient for their inversion.

If $\lambda_k$ is an eigenvalue of $S$ then $\lambda_k /(1-\lambda_k^4)$ is an
eigenvalue of $Q$. If $\lambda_k$ is a small eigenvalue, then $\left|
  \lambda_k^4 \right| \lll 1$. In this case using $\lambda_k
/(1-\lambda_k^4)$ for the calculations instead of $\lambda_k$ does not make
any difference. If $\lambda_k$ is a large eigenvalue of $S$, then $\lambda_k
/(1-\lambda_k^4) \approx -1/\lambda_k^3$, which is negligible compared to the
relevant small eigenvalues. That is, the relevant eigenvalues of $S$ become
the largest eigenvalues of $Q$.

\begin{figure}
\begin{center}
\resizebox{!}{\figheight}{\includegraphics{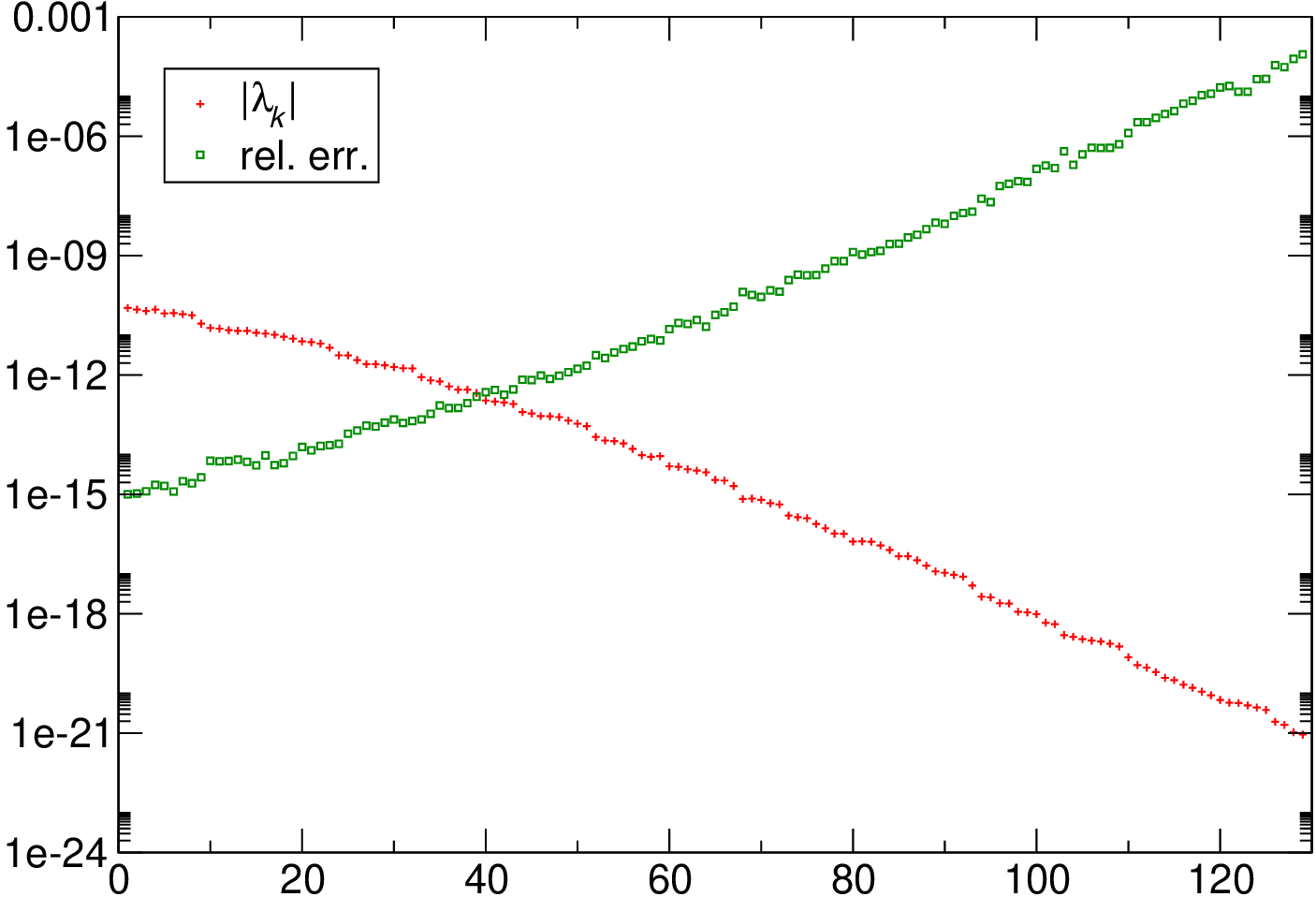}}
\end{center}
\caption{The absolute values of the relevant eigenvalues found by the double
  precision version of ARPACK (red crosses) and their relative errors (green
  squares) on a typical lattice of size $6^3\times 100$.}
\label{fig:arpack}
\end{figure}

The procedure for finding the relevant eigenvalues was as follows. After
fixing the temporal gauge the matrices $B_k$ were built. Then from
(\ref{eqn:S}) and (\ref{eqn:Sinv}) the matrices $S$ and $S^{-1}$ were
constructed. Since $S$ and $S^{-1}$ are very badly conditioned, their
construction as well as working with them requires high precision. For these
and the latter calculations the GNU multiple precision arithmetic library (GNU
MP) was used. Then after inverting $S+S^{-1}$ and $S-S^{-1}$ the largest
several eigenvalues of $Q$ (the relevant ones) were obtained using the double
precision version of ARPACK. This last step may sound doubious but in fact the
double ARPACK was found to be able to reliably find the eigenvalues that were
not more than 10 orders of magnitude smaller than the largest one (see Figure
\ref{fig:arpack}).

\section{Consequences of the {\boldmath$Z_3$} symmetry}
\label{sec:z3}

\subsection{Consequences for \boldmath{}$Z_{N_1,\dots,N_{n_s}}$}

From (\ref{eqn:detdet2}) it can be seen that the quantity within the
expectation value signs in equation (\ref{eqn:zmulat3}) is periodic in each
$\hat{\mu}_i$ with a periodicity of $2\pi / L_t$ configuration by
configuration. Therefore, the lattice grand canonical partition function
(\ref{eqn:zmulat1}) is also periodic with $2\pi/L_t$ in each $\hat{\mu}_i$.

Performing a $Z_3$ transformation, that is, multiplying all the time-like
links on the last timeslice of an $SU(3)$ configuration $U$ by
$\varepsilon_j$ ($\varepsilon_j= e^{2\pi i\cdot j/3}$, $j=0,1,2$) gives
another $SU(3)$ configuration denoted by $U^{\varepsilon_j}$. Then the
partition function can be written as
\begin{equation}
  Z(i\hat{\mu}_1,\dots,i\hat{\mu}_{n_s}) = \frac13 \sum_{j=0}^2
  \int[\mathrm{d}U^{\varepsilon_j}]\,e^{-S_g[U^{\varepsilon_j}]}
\,\prod_{i=1}^{n_s} \det
  M(m_i,i\hat{\mu}_i,U^{\varepsilon_j})^{n_i/4}.
\label{eqn:zmulatz3_1}
\end{equation}
The functional measure and the gauge action are both symmetric with respect to
$Z_3$ transformations \cite{Weiss:1980rj}. Thus,
\newlength{\ZMM}
\begin{eqnarray}
  Z(i\hat{\mu}_1,\dots,i\hat{\mu}_{n_s}) &=& \int[\mathrm{d}U]\,e^{-S_g[U]}
  \,\frac13 \sum_{j=0}^2 \prod_{i=1}^{n_s} \det
  M(m_i,i\hat{\mu}_i,U^{\varepsilon_j})^{n_i/4}\nonumber\\
  \label{eqn:zmulatz3_2}
    &=& \int[\mathrm{d}U]\,e^{-S_g[U]}\,\prod_{i=1}^{n_s}\det
    M(m_i,0,U)^{n_i/4}
 \times \frac13 \sum_{j=0}^2 \prod_{i=1}^{n_s} \left( \frac{\det
      M(m_i,i\hat{\mu}_i,U^{\varepsilon_j})}{\det M(m_i,0,U)} \right)^{n_i/4}\\
  &=& Z \cdot \left< \frac13 \sum_{j=0}^2 \prod_{i=1}^{n_s} \left( \frac{\det
          M(m_i,i\hat{\mu}_i,U^{\varepsilon_j})}{\det 
          M(m_i,0,U)} \right)^{n_i/4} \right>.\nonumber
\end{eqnarray}

Since the $U \to U^{\varepsilon_j}$ transformation can be applied in
eq.~(\ref{eqn:det1}) by simply multiplying $S$ by $\varepsilon_j$, the ratios
of the determinants in (\ref{eqn:zmulatz3_2}) can be rewritten as
\begin{equation}
\begin{split}
\frac{\det M(m_i,i\hat{\mu}_i,U^{\varepsilon_j})}{\det M(m_i,0,U)} &= 
 \prod_{k=1}^{3V} \left| \frac{1 - \lambda_k^{(i)} \varepsilon_j
 \,e^{i\hat{\mu}_i L_t}}{ 1 - \lambda_k^{(i)}}  \right|^2 =
 \prod_{k=1}^{3V} \left| \frac{1 - \lambda_k^{(i)} 
 \,e^{i\hat{\mu}_i L_t + i \frac{2\pi}{3}j}}{ 1 - \lambda_k^{(i)}} \right|^2\\
&=\frac{\det M\!\left(m_i,i\hat{\mu}_i + i\frac{2\pi}{3L_t}
    ,U\right)}{\det M(m_i,0,U)}.
\label{eqn:detz3_1}
\end{split}
\end{equation}
Combining (\ref{eqn:detz3_1}) with (\ref{eqn:zmulatz3_2}) we obtain
\begin{multline}
Z(i\hat{\mu}_1,\dots,i\hat{\mu}_{n_s}) = \frac13 \Bigg[
  Z(i\hat{\mu}_1,\dots,i\hat{\mu}_{n_s}) +
  Z\!\left(i\hat{\mu}_1 + i\frac{2\pi}{3L_t},\dots,i\hat{\mu}_{n_s}+
  i\frac{2\pi}{3L_t}\right) \\
  + Z\!\left(i\hat{\mu}_1 + i\frac{4\pi}{3L_t},\dots,i\hat{\mu}_{n_s}+
  i\frac{4\pi}{3L_t}\right) \Bigg],
\end{multline}
which means that the grand canonical partition function has an extra
periodicity: if $2\pi i/3L_t$ is added to all the chemical potentials then the
value of the partition function remains unchanged \cite{Roberge:1986mm}.
\begin{equation}
  Z(i\hat{\mu}_1,\dots,i\hat{\mu}_{n_s}) = 
  Z\!\left(i\hat{\mu}_1 + i\frac{2\pi}{3L_t},\dots,i\hat{\mu}_{n_s}+
    i\frac{2\pi}{3L_t}\right)
\end{equation}
Therefore, the canonical partition functions $Z_{N_1,\dots,N_{n_s}}$ where
the total number of quarks $N_1+\cdots+N_{n_s}$ is not divisible by 3 are
zero \cite{Kratochvila:2006jx}.

Taking this into account the expectation value of the first order expansion in
(\ref{eqn:1stexp}) gives
\begin{equation}
  \left< \left( \frac{\det M(i\hat{\mu})}{\det M(0)} \right)^{n_t/4}  \right>
  \approx \left< 1 + \frac{n_t}{4} \sumi\li + \frac{n_t}{4} \sumi\lis \right>.
\label{eqn:1stexpexp}
\end{equation}

\subsection{Application on a term by term basis}

Let $A[U]$ be a gauge invariant quantity (a gauge invariant function of the
gauge configuration $U$). Then the expectation value of $A[U]$ is
\begin{equation}
  \left< A[U] \right> = \frac{1}{Z} \int[\mathrm{d}U]\,e^{-S_g[U]}
  \det M(m_1,0,U)^{n_1/4}\,\cdots \, \det M(m_{n_s},0,U)^{n_{n_s}/4} \cdot
  A[U]. 
\label{eqn:Aexp1}
\end{equation}
Using the $Z_3$ invariance of the gauge action and the integration measure one
can rewrite (\ref{eqn:Aexp1}) as
\begin{equation}
  \left< A[U] \right> = \frac{1}{Z} \int[\mathrm{d}U]\,e^{-S_g[U]}
  \prod_{i=1}^{n_s} \det M(m_i,0,U)^{n_i/4}
\times \frac13 \sum_{j=0}^2 A\!\left[U^{\varepsilon_j} \right] \cdot
 \prod_{i=1}^{n_s} \left( \frac{\det
      M(m_i,0,U^{\varepsilon_j})}{\det M(m_i,0,U)} \right)^{n_i/4},
\end{equation}
that is,
\begin{equation}
\left< A[U] \right> = \left< \frac13 \sum_{j=0}^2 A\!\left[U^{\varepsilon_j}
 \right] \cdot  \prod_{i=1}^{n_s} \left( \frac{\det
      M(m_i,0,U^{\varepsilon_j})}{\det M(m_i,0,U)} \right)^{n_i/4} \right>.
\label{eqn:Aexpz3}
\end{equation}
Using (\ref{eqn:detz3_1}) an expansion similar to (\ref{eqn:1stexp}) can be
applied to these determinant ratios.

All the quantities of the form of (\ref{eqn:expvar}) are gauge independent,
therefore, each term of the series expansion can individually be taken as
$A[U]$. This way the $Z_3$ symmetric form of (\ref{eqn:Aexpz3}) can be applied
to each term in the series expansion.  As an example, if we have only one
staggered field ($n_s=1$) with number of tastes $n_t$ then the expectation
value of the first order terms in eq.~(\ref{eqn:1stexp}) up to leading order
become
\begin{equation}
\left< \sumi\li \right> \stackrel{\text{LO}}{=} 
\left< \sumi\lis \right> \stackrel{\text{LO}}{=} 
\left< -\frac{n_t}{4} \left| \sumi\li \right| ^2 \right>.
\end{equation}
Applying this technique term by term the series expansion of
(\ref{eqn:1stexp}) and (\ref{eqn:1stexpexp}) up to third order becomes
\begin{multline}
  \left< \left( \frac{\det M(i\hat{\mu})}{\det M(0)} \right)^{n_t/4}  \right> 
\approx \Bigg< 1 +  \frac{n_t}{12}\sumi\li^3
  -\frac{n_t^2}{32}\left(\sumi\li\right) \left(\sumi\li^2\right) +
  \frac{n_t^3}{384} \left(\sumi\li\right)^3 \\
+  \frac{n_t}{12}\sumi\lis^3
  -\frac{n_t^2}{32}\left(\sumi\lis\right) \left(\sumi\lis^2\right) +
  \frac{n_t^3}{384} \left(\sumi\lis\right)^3 \Bigg>\\
+ e^{3i\hat{\mu}L_t} \cdot \Bigg< -  \frac{n_t}{12}\sumi\li^3
  +\frac{n_t^2}{32}\left(\sumi\li\right) \left(\sumi\li^2\right) -
  \frac{n_t^3}{384} \left(\sumi\li\right)^3 \Bigg> \\
+ e^{-3i\hat{\mu}L_t} \cdot \Bigg< -  \frac{n_t}{12}\sumi\lis^3
  +\frac{n_t^2}{32}\left(\sumi\lis\right) \left(\sumi\lis^2\right) -
  \frac{n_t^3}{384} \left(\sumi\lis\right)^3 \Bigg>.
\label{eqn:3rdexpexp}
\end{multline}

In the third order expansion in (\ref{eqn:3rdexpexp}) all the terms are
already $Z_3$ invariant.  If a term in the series expansion is $Z_3$ invariant
then it does not change when the procedure of (\ref{eqn:Aexpz3}) is applied to
it. When the procedure (\ref{eqn:Aexpz3}) is applied to a non--$Z_3$ invariant
term, its expectation value becomes the expectation value of the sum of higher
order terms. This procedure can be continued order by order and as a result,
all the remaining terms in all the quark number sectors of the series
expansion of $Z(i\hat{\mu}_1,\dots,i\hat{\mu}_{n_s})$ become $Z_3$ invariant.

\section{Application to baryons}
\label{sec:baryon}

In principle the method described in Sections \ref{sec:cpf_mass} and
\ref{sec:cpf_lattice} can be used to measure the mass of the lowest state in
any quark number sector. For example, one can think of the di-baryon
(deuteron), or the bound states of even more baryons. However, in most
cases technical difficulties occur. Let us examine the case when one tries to
measure the mass of a baryon, for example the proton. For that we use two
staggered quark fields, one for the $u$ quark with $n_u$ tastes and one for
the $d$ quark with $n_d$ tastes. (We omitted the third light quark, the $s$
quark, the inclusion of which in our case does not change the picture
significantly.)

The proton is believed to be the lowest state in the $N_u=2,N_d=1$ channel,
therefore, according to (\ref{eqn:fdifflin}) we need to examine the low
temperature behaviour of
\begin{equation}
F_{N_u=2,N_d=1}(T) - F_{N_u=0,N_d=0}(T) = -T \ln \left(
  \frac{Z_{2,1}(T)}{Z_{0,0}(T)} \right).
\end{equation}
According to (\ref{eqn:znlo4})--(\ref{eqn:expreplace}) the relevant canonical
partition functions to leading order are
\begin{equation}
\begin{split}
Z_{2,1} &\stackrel{\text{LO}}{=} Z \cdot \left< \frac{n_un_d}{32} \left(
    \sum_{k=1}^{3V} {\lambda_k^{(u)}}^2 \right) \left( \sum_{k=1}^{3V}
    \lambda_k^{(d)}\right) - \frac{n_u^2n_d}{128}\left(
    \sum_{k=1}^{3V} \lambda_k^{(u)} \right)^2 \left( \sum_{k=1}^{3V}
    \lambda_k^{(d)}\right) \right>,\\
Z_{0,0} & \stackrel{\text{LO}}{=} Z \cdot \left< 1 \right>,
\end{split}
\end{equation}
therefore, for the proton mass one obtains
\begin{equation}
  am_p = \lim_{L_t \to \infty} - \frac{1}{L_t} \ln \left<
    \frac{n_un_d}{32} \left( \sum_{k=1}^{3V} {\lambda_k^{(u)}}^2 \right)
    \left( \sum_{k=1}^{3V}  \lambda_k^{(d)}\right) 
  - \frac{n_u^2n_d}{128}\left( \sum_{k=1}^{3V} \lambda_k^{(u)} \right)^2
    \left( \sum_{k=1}^{3V}\lambda_k^{(d)}\right) \right>.
\label{eqn:amp}
\end{equation}
As the temperature decreases ($L_t$ increases) the eigenvalues become smaller
and smaller, and only the leading order term matters in the limit.

The formulae for the masses of the 2-baryon, 3-baryon, etc.~channels can be
obtained similarly. These can in principle be used to measure the bonding
energy of several-baryon states.

The description seems simple, but there is one difficulty. The expression of
which the expectation value is taken in equation (\ref{eqn:amp}) can be any
complex number, whose real part can be both positive and negative. Its
expectation value is much smaller than its value at a typical gauge
configuration. On a $6^3\times 24$ staggered lattice with
$a\approx 0.33\,\text{fm}$, $m_{\pi} \approx 330 \,\text{MeV}$ and $T \approx
25\,\text{MeV}$ this value at a typical gauge configuration is of $O(10^{-10})$
while the expected order of magnitude of the expectation value is
$O(10^{-20})$. That means that the number of configurations needed for a
correct result would be of $O(10^{20})$, which is prohibitive. The problem
becomes even more severe when one decreases the temperature in order to get
closer to the $T\to 0$ limit.

\section{Application to mesons}
\label{sec:isospin}

When $n_d=n_u=n_t/2$, $m_d=m_u$ and we are looking at one of the $N_d=-N_u$
sectors this sign problem does not arise. These sectors can be labelled with
one parameter, the third component of the isospin $I_3 = (N_u-N_d)/2$. Since
$\lambda_k^{(u)} = \lambda_k^{(d)}$ for all $k$, we will write $\lambda_k$
only. 

The lowest state in the $I_3=1$ sector is expected to be the Goldstone
pion. Its partition function can be written as the expectation value
\begin{equation}
  Z_{I_3=1} \stackrel{\text{LO}}{=} Z_{N_u=1,N_d=-1} \stackrel{\text{LO}}{=}
  Z\cdot \left< \frac{n_t^2}{64} \left| \sumi\li \right|^2 \right>,
\label{eqn:zpion}
\end{equation}
which is a manifestly positive polynomial of the eigenvalues. Therefore, it
can be easily evaluated, and by taking the zero temperature limit 
\begin{equation}
  am_{I_3=1,\pi} = \lim_{L_t \to \infty} - \frac{1}{L_t} \ln \left<
  \frac{n_t^2}{64} \left| \sumi\li \right|^2 \right> 
  \label{eqn:ampi}
\end{equation}
one directly obtains the mass of the lowest state in the $I_3=1$ channel.

The formulae for the energies of the lowest state in higher $I_3$ channels can
be obtained similarly. These can be used to investigate pion-pion scattering
and several-pion states.

The result for the pion mass given in equation (\ref{eqn:ampi}) obtained using
purely thermodynamic considerations can be compared to formula
(\ref{eqn:gibbs}). If $L_t$ is large, then after taking the logarithm the
factor $n_t^2/64$ gives a negligible contribution compared to that of the
sum. If the temporal extension is large even compared to the spatial volume,
then the sum in (\ref{eqn:ampi}) is dominated by the largest of the small
eigenvalues. In this case equations (\ref{eqn:ampi}) and (\ref{eqn:gibbs})
evaluated on a single configuration yield approximately the same
results. However, while Ref.~\cite{Gibbs:1986hi} only states that relation
(\ref{eqn:gibbs}) holds configuration by configuration and does not mention
how to obtain results over an ensemble of configurations, equation
(\ref{eqn:ampi}) describes a method for taking the ensemble average.

\section{Results}
\label{sec:results}

\subsection{Dynamical staggered fermions}

We performed calculations using dynamical staggered configurations to measure
the masses in the first isospin channel as described in Section
\ref{sec:isospin}. We used the Wilson plaquette action for the gauge fields
and unimproved staggered fermion action. In order to be able to check whether
the root taking of the fermion determinant changes the results significantly,
calculations were done using rooted staggered fermions with $n_t=2$
($n_u=n_d=1$) and $n_t=4$ ($n_u=n_d=2$) as well as unrooted fermions with
$n_t=8$ ($n_u=n_d=4$).

\begin{table}
\begin{center}
\begin{tabular}{c|ccc}
\hline \hline
$1/aT$ & \multicolumn{3}{c}{Number of configurations} \\
 & $n_t=2$ & $n_t=4$ & $n_t=8$ \\ \hline
50 & 331 & 322 & -- \\
100 & 1196 & 935 & 701 \\
200 & 323 & 605 & 467 \\
300 & 168 & 255 & -- \\ \hline \hline
\end{tabular}
\end{center}
\caption{The number of configurations used for dynamical staggered
  calculations with a spatial volume of $6^3$.}
\label{tab:dyn_6_num}
\end{table}

For the $n_t=2$ runs the gauge coupling was $\beta=4.8$. The lattice spacing
was $a=0.41\,\text{fm}$, measured from the string tension $\sigma$ using the
value of $\sqrt{\sigma}= 465\,\text{MeV}$ \cite{Edwards:1997xf}. For the
$n_t=4$ case $\beta=4.3$ and $a=0.42\,\text{fm}$ and for the $n_t=8$ case
$\beta=3.8$ and $a=0.44\,\text{fm}$. In all three cases the bare quark mass
was $am_q=0.04$ and the spatial extension of the lattice was $L_s=6$.  In the
two rooted case we used temporal lattice extensions of $L_t=50,100,200,300$
while in the unrooted case only $L_t=100,200$ was used. Table
\ref{tab:dyn_6_num} contains the number of configurations for each setup.

\begin{table}
\begin{center}
\begin{tabular}{c|ccc}
\hline \hline
$1/aT$ & \multicolumn{3}{c}{$aF_{I_3=1}-aF_{I_3=0}$} \\
& $n_t=2$ & $n_t=4$ & $n_t=8$ \\ \hline
50 & 0.5344(12) & 0.4971(12) & -- \\
100 & 0.5066(2) & 0.4826(4) & 0.4639(4) \\
200 & 0.4931(2) & 0.4760(1) & 0.4641(3) \\
300 & 0.4876(3) & 0.4730(3) & -- \\ \hline
$\to \infty$ & 0.4787(3) & 0.4688(3) & 0.4643(7) \\ \hline
$am_{\pi,\text{sp}}$ & 0.47864(3) & 0.46903(4) & 0.46426(3) \\ \hline\hline 
\end{tabular}
\end{center}
\caption{The differences of the free energies, their $T\to 0$ extrapolated
  values and the spectroscopic pion masses on dynamical staggered
  configurations with a spatial volume of $6^3$.}
\label{tab:dyn_6}
\end{table}

\begin{figure}
\begin{center}
\resizebox{!}{\figheight}{\includegraphics{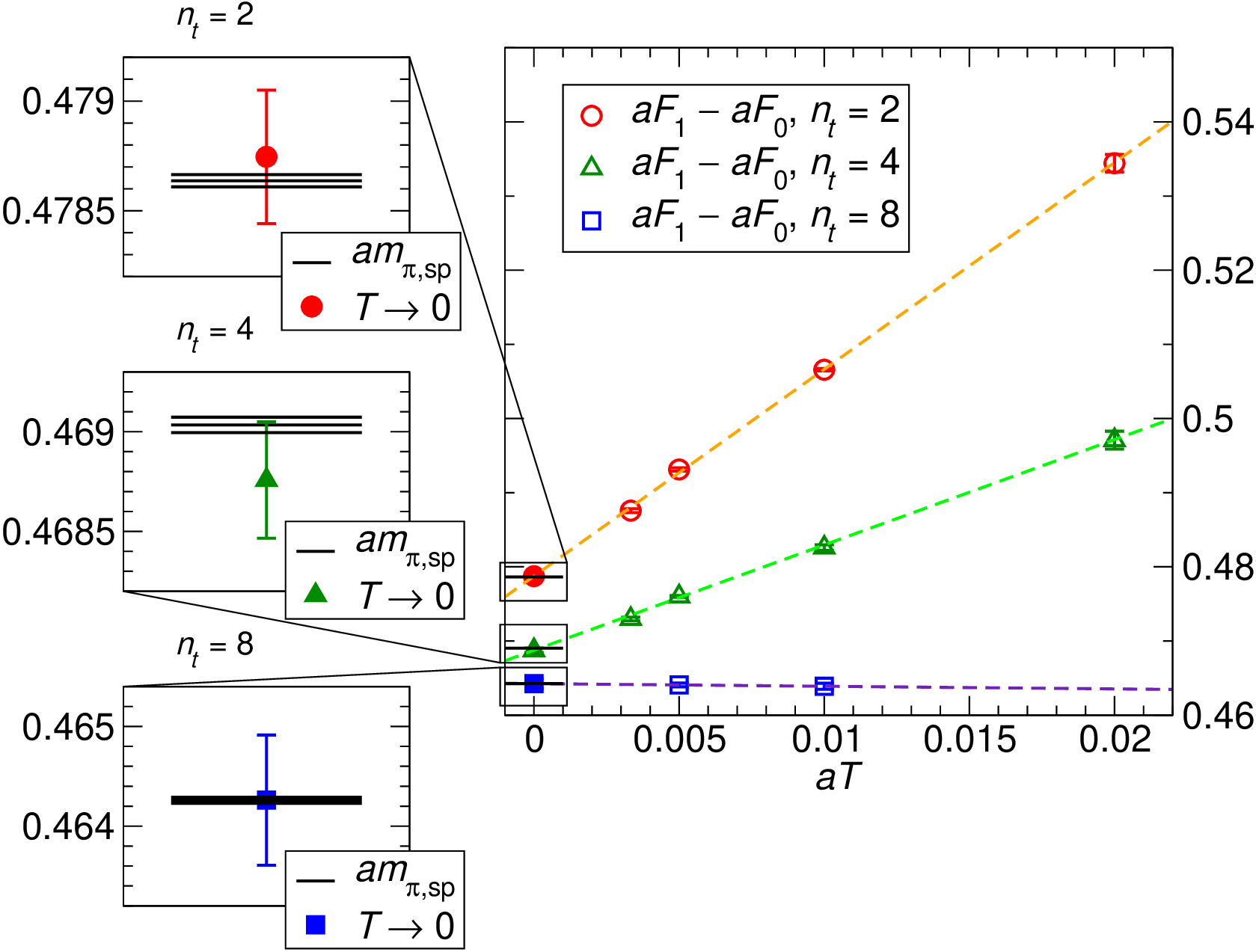}}
\end{center}
\caption{The differences of the free energies of the isospin one and isospin
  zero sectors as a function of the temperature on dynamical staggered
  configurations with a spatial volume of $6^3$. The dashed lines show the
  linear fits to the data points. The $T\to 0$ extrapolated values are
  compared to the spectroscopic pion masses.}
\label{fig:dyn_6_1pion}
\end{figure}

Using equation (\ref{eqn:zpion}) the difference of the free energies
$aF_{I_3=1}-aF_{I_3=0}$ were measured on each set of configurations. These are
listed in Table \ref{tab:dyn_6}. According to equation (\ref{eqn:fdifflin})
the mass of the ground state in the $I_3=1$ channel can be obtained using a
linear extrapolation to $T=0$. For comparison we measured the pion mass in all
cases using the ordinary spectroscopic method, which will be denoted by
$m_{\pi,\text{sp}}$. The measured free energy values, the linear fits and the
comparisons to the spectroscopic pion masses can be seen in Figure
\ref{fig:dyn_6_1pion}.

As can be seen from Figure \ref{fig:dyn_6_1pion} the mass of the ground state
in the $I_3=1$ sector agrees with the spectroscopic pion mass within
errorbars for both rooted and unrooted staggered fermions.

\subsection{Quenched case}

Equation (\ref{eqn:ampi}) can be rewritten as
\begin{equation}
  am_{I_3=1,\pi} = \lim_{L_t \to \infty} \left[ - \frac{1}{L_t} \ln
    \left( \frac{n_t^2}{64} \right) - \frac{1}{L_t} \ln \left< \left| \sumi \li
      \right|^2 \right> \right] = \lim_{L_t \to \infty} - \frac{1}{L_t} \ln
  \left< \left| \sumi \li     \right|^2 \right>.
\label{eqn:qmpi}
\end{equation}
The r.h.s.~of equation (\ref{eqn:qmpi}) does not explicitly contain the number
of staggered tastes. The quantity
\begin{equation}
\text{``}aF_{I_3=1}-aF_{I_3=0}\text{''} = - \frac{1}{L_t} \ln \left< \left|
    \sumi \li     \right|^2 \right>
\label{eqn:qfdiff}
\end{equation}
can be evaluated on quenched configurations as well. The question arises
naturally: If one measures the pion mass on a quenched ensemble using regular
staggered spectroscopy and evaluates the expression in (\ref{eqn:qmpi}) with
the same fermion mass, will these be the same?

\begin{figure}
\begin{center}
\resizebox{!}{\figheight}{\includegraphics{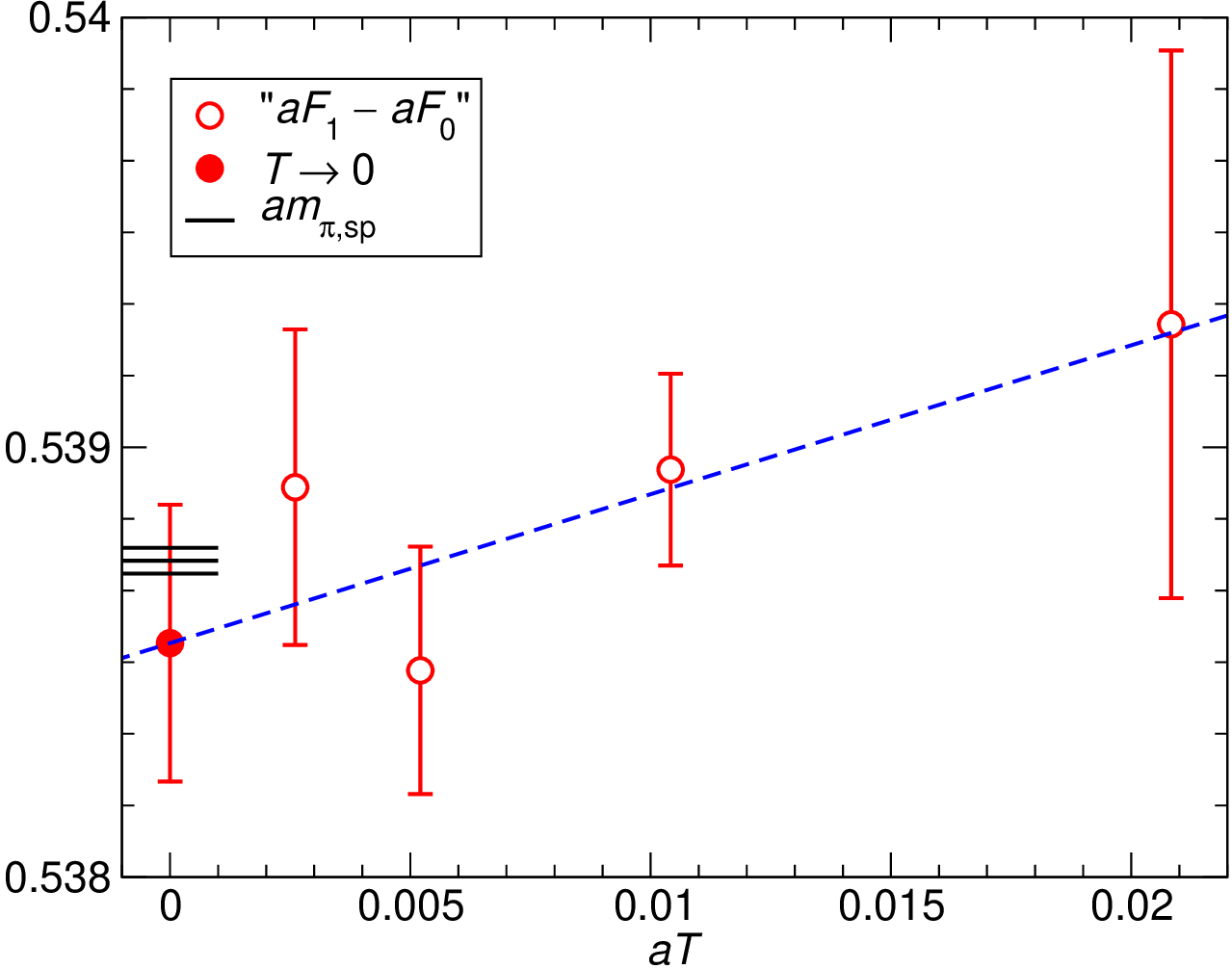}}
\end{center}
\caption{The ``differences of the free energies'' of the isospin one and
  isospin zero sectors as a function of the temperature on quenched
  configurations with a spatial volume of $6^3$ and bare quark mass
  $am_q=0.04$. The dashed line shows the linear fit to the data points. The
  $T\to 0$ extrapolated value is compared to the spectroscopic pion mass.}
\label{fig:quenched}
\end{figure}

To find this out we performed calculations on quenched configurations
generated using the Wilson plaquette gauge action. The spatial extension of
the lattice was $L_s=6$, the gauge coupling was $\beta=5.6$ and the
corresponding lattice spacing was $a=0.21\,\text{fm}$
\cite{Edwards:1997xf}. The time extension of the used lattices were
$L_t=48,96,192,384$ and for the measurements we used a bare quark mass of
$am_q=0.04$. The number of configurations used are listed in Table
\ref{tab:quenched_num}. The results are summarized in Table \ref{tab:quenched}
and the linear extrapolation is shown in Figure \ref{fig:quenched}. The
comparison shows that the result obtained from the free energies is
consistent with the spectroscopic pion mass.

\begin{table}
\begin{center}
\begin{tabular}{c|c}
\hline \hline
$1/aT$ & Number of configurations \\ \hline
48 & 2502  \\
96 & 1852  \\
192 & 731  \\
384 & 412  \\ \hline \hline
\end{tabular}
\end{center}
\caption{Number of quenched configurations.}
\label{tab:quenched_num}
\end{table}

The partition function contains all the information about the degrees of
freedom present in the system, therefore, the free energy should be able to
make a difference between dynamical and quenched configurations. The results,
however, show that both types of ensembles yield a free energy that is
consistent with particles of mass equal to the spectroscopic pion mass
present in the system. Thus, one cannot tell this way whether a given set of
configurations is from a dynamical or a quenched ensemble.

\begin{table}
\begin{center}
\begin{tabular}{c|cccc}
\hline \hline
$1/aT$ & ``$aF_{I_3=1}-aF_{I_3=0}$'' \\ \hline
48 & 0.5393(6) \\
96 & 0.5389(2) \\
192 & 0.5385(3) \\
384 & 0.5389(4) \\ \hline
$\to \infty$ & 0.5385(3) \\ \hline
$am_{\pi,\text{sp}}$ & 0.53874(3) \\ 
\hline  \hline 
\end{tabular}
\end{center}
\caption{The ``differences of the free energies'' of the isospin one and
  isospin zero sectors, their $T\to 0$ extrapolated value and the
  spectroscopic pion mass on quenched configurations.}
\label{tab:quenched}
\end{table}

\section{Conclusions}
\label{sec:conclusions}

We have proposed a spectroscopic method based on purely thermodynamical
considerations. The formulae obtained show the relation between the
eigenvalues of the reduced staggered fermion matrix and the hadron
spectrum. The method not only clarifies the findings of
Ref.~\cite{Gibbs:1986hi} in connection with the Goldstone pion mass, but also
extends them.  In principle, the method can be used to obtain the mass of the
lightest particle in a given quark number sector. For example, in principle,
the mass of the di-baryon could be obtained. However, it turns out that the
application even to one-baryon states is computationally very
demanding. Nevertheless, we successfully applied our method to the Goldstone
pion. In the calculations presented the mass of the lowest state in the
$I_3=1$ sector is in agreement with the pion mass obtained using the ordinary
spectroscopic method. This indicates that the method presented in Section
\ref{sec:isospin} is a valid way of finding the pion mass.

\section*{Acknowledgements}
We would like to thank G.~I.~Egri, C.~Hoelbling and S.~D.~Katz for their
help. This research was partially supported by OTKA Hungarian Science Grants
No.\ T34980, T37615, M37071, T032501, AT049652 and by DFG German Research
Grant No.\ FO 502/1-1.  The computations were carried out on the 370 processor
PC cluster of E\"otv\"os University and on the 1024 processor PC cluster of
Wuppertal University. We used a modified version of the publicly available
MILC code \cite{MilcCode} with next-neighbor communication architecture for
PC-clusters \cite{Fodor:2002zi}.

\bibliographystyle{spectr_cpf}
\bibliography{spectr_cpf}

\end{document}